\documentclass[12pt,a4paper]{article}

\usepackage[left=5em]{geometry}
\usepackage{amsmath}

\title{Inflation and New Agegraphic Dark Energy}
\author{Cheng-Yi Sun\footnote{cysun@mailis.gucas.ac.cn; ddscy@163.com}\
$^{,1}$\ and Rui-Hong Yue\footnote{yueruihong@nbu.edu.cn}\ $^{,2}$
\\
 {$^1$\small Institute of Modern Physics, Northwest University, Xian 710069, P.R.
 China.}\\
{$^2$\small Faculty of Science, Ningbo University, Ningbo 315211,
P.R. China.}}

\begin{document}
\maketitle
\begin{abstract}
In the note, we extend the discussion of the new agegraphic dark
energy (NADE) model to include the inflation stage. Usually, in the
inflation models, for convenience the conformal time $\eta$ is set
to be zero at the end of inflation. This is incompatible with the
NADE model since $\eta=0$ indicates the divergence of NADE. To avoid
the difficulty, we can redefine the conformal time as $\eta+\delta$.
However, we find that the positive constant $\delta$ must be so
large that NADE can not become dominated at present time.
\end{abstract}



Increasing evidence suggests that the expansion of our universe is
being accelerated \cite{Supernova,WMAP,LSS}. Within the framework of
the general relativity, the acceleration can be phenomenally
attributed to the existence of a mysterious exotic component with
negative pressure, namely the dark energy \cite{dark energy1,dark
energy2}. However, we know little about the nature of dark energy.
The most nature, simple and important candidate for dark energy is
the Einstein's cosmological constant, which can fit the observations
well so far. But the cosmological constant is plagued with the
well-known fine-tuning and cosmic coincidence problems \cite{dark
energy1,dark energy2}. Dark energy has become one of the most active
fields in the modern cosmology.

Recently, a new dynamical model of dark energy, the new agegraphic
dark energy (NADE) model, is proposed \cite{0708.0884}. The energy
density of agegraphic dark energy is proposed to be \cite{0708.0884}
\begin{equation}
  \label{NADE}
  \rho_q=\frac{3n^2M_p^2}{\eta^2}.
\end{equation}
Here $n$ is a constant pramameter, $M_p=(8\pi G)^{-1/2}$ and $\eta$
is the conformal age of the universe
\begin{equation}
  \label{eta}
  \eta=\int^t_0{\frac{dt'}{a(t')}}=\int^a_0{\frac{da}{Ha^2}},
\end{equation}
where $a$ is the scale factor, $H\equiv\dot{a}/a$ is the Hubble
parameter and a dot denotes the derivative with respect to the
cosmic time $t$.

The NADE model is successful in explaining the accelerated expansion
and in fitting the observational data \cite{0904.0928,1011.6122}.
The evolution of NADE has been surveyed in detail. It has been shown
that NADE becomes dominated in the future, and negligible in the
matter-dominated epoch and in the radiation-dominated epoch, as
expected \cite{0708.0884}.

Since it is generally believed that before the radiation-dominated
epoch there exists another important stage, \emph{Inflation}, it is
natural for us to extend the discussion of the NADE model to include
the inflation stage. It seems that only trivial results would be
obtained since NADE should be negligible in the inflation stage.
However, we find that there exists incompatibilities between the
inflation scenario and the NADE model.

Let us show it. From Eq.(\ref{NADE}), we know that once the
conformal time is given, $\rho_\eta$ will be determined. It is known
that in the inflation models the conformal time $\eta$ is an
important parameter which is widely used in calculating the
primordial spectrum of the curvature perturbation, and usually is
defined as
\begin{equation}
  \label{deta}
  d\eta=\frac{dt}{a}=\frac{da}{Ha^2}.
\end{equation}
Since in a slow-rolling inflation model, $H$ is constant
approximately, then by integrating Eq.(\ref{deta}), we can get
\begin{equation}
  \label{etaH}
  \eta-\eta_i\simeq\frac{1}{H}\Big(\frac{1}{a_i}-\frac{1}{a}\Big)\simeq-\frac{1}{Ha}-\Big(-\frac{1}{H_ia_i}\Big),
\end{equation}
where the subscript $i$ denotes the value of the corresponding
parameter at some initial moment during the inflation stage. Then by
choosing $\eta_i=-\frac{1}{H_ia_i}$, we can get a well-known result
\begin{equation}
  \label{etaInflation}
  \eta\simeq-\frac{1}{Ha}.
\end{equation}
Of course, we can express $\eta$ more precisely by using the
slow-rolling parameters. Here what is important is not the precise
form but the fact that $\eta$ is \emph{negative} in the inflation
scenario. In fact, it is well known that, during the inflation
stage,
\[
  -\infty<\eta<0,
\]
and usually, in the literature it is set
\[
  \eta_E=0
\]
at the end of slow-roll inflation for convenience \cite{9807278}.
Hereafter the subscript $E$ denotes the value of the corresponding
parameter at the end of the inflation. On the other hand, in the
radiation-dominated epoch, with $\eta_E=0$ and Eq.(\ref{eta}) we get
\begin{equation}
  \label{etaRD}
  \eta=\frac{1}{Ha}>0.
\end{equation}
Then the fact that $\eta$ is negative during the inflation stage and
positive in the radiation-dominated epoch indicates that $\eta$ must
be zero at some moment before the radiation-dominated epoch.
Obviously, this is unacceptable since $\eta=0$ indicates that
$\rho_q$ defined in Eq.(\ref{NADE}) is infinite. So
Eq.(\ref{etaInflation}) that is widely used in the inflation models
contradicts the NADE model.

It seems that the contradiction can be removed easily if we do not
choose $\eta_i=-\frac{1}{a_iH_i}$. Then we can define a positive
conformal time from Eq.(\ref{etaH}) as
\begin{equation}
  \label{etaInflDelta}
  \eta\simeq\delta-\frac{1}{Ha},
\end{equation}
where $\delta=\eta_i+\frac{1}{a_iH_i}$ is a constant parameter. To
remove the zero point of $\eta$, we should require that
\[
  \delta>\frac{1}{Ha}
\]
holds during the inflation stage. This can be guaranteed by the
requirement
\begin{equation}
  \label{delta}
  \delta\ge\frac{1}{H_Ia_I},
\end{equation}
since $Ha$ is increasing in the inflation stage. Hereafter the
subscript $I$ denotes the value of the corresponding parameter at
the beginning of the inflation. Then at the end of the inflation, we
have
\begin{equation}
  \label{etaE}
  \eta_E=\delta-\frac{1}{a_EH_E}\ge\frac{1}{a_IH_I}-\frac{1}{a_EH_E}.
\end{equation}
In fact, the definition of Eq.(\ref{etaInflation}) is to set
$\eta_E=0$, while Eq.(\ref{etaInflDelta}) is to set
\begin{equation}
 \nonumber
 \eta_I\ge0.
\end{equation}
Then, with Eq.(\ref{etaInflDelta}), the contradiction
between inflation and NADE displayed in the last paragraph is
eliminated.

However, the other problem results from Eq.(\ref{etaInflDelta}).
From Eqs.(\ref{eta}), after the end of the inflation, we have
\begin{equation}
  \label{etaAfterInf}
  \eta=\eta_E+\int_{t_E}^t\frac{dt'}{a(t')}.
\end{equation}
Obviously, at the present time, we have
\begin{equation}
  \label{eta0GDelta}
  \eta_0>\eta_E\ge\frac{1}{a_IH_I}-\frac{1}{a_EH_E},
\end{equation}
where Eq.(\ref{etaE}) has been used. Hereafter the subscript $0$
denotes the value of the corresponding parameter at the present
time. Then by defining $\Omega_q=\frac{n^2}{H^2\eta^2}$, we have
\begin{equation}
  \label{OmegaqLess}
  \Omega_{q0}<\frac{n^2}{H_0^2}a_I^2H_I^2\Big(1-\frac{a_IH_I}{a_EH_E}\Big)^{-2}.
\end{equation}
We define the comoving wavenumber $k_*$ corresponding to the present
Hubble scale as
\begin{equation}
  \label{k}
    k_*=H_0a_0.
\end{equation}
Then to solve the horizon problem, we should require
\[
  k_*>H_Ia_I.
\]
In fact, the requirement is not enough. Generally, in order for the
cosmological perturbation on the Hubble scale to be generated from
the quantum fluctuation, we should require
\begin{equation}
  \label{kggHI}k_*\gg H_Ia_I.
\end{equation}
Then together with Eq.(\ref{k}), we have
\begin{equation}
  \label{HIH0}\frac{H_I^2a_I^2}{H_0^2a_0^2}\ll1
\end{equation}
By analyzing the observational data, it has been shown in
\cite{1011.6122} that with the choice of $a_0=1$,
$n=2.886^{+0.084}_{-0.082}$ at $1\sigma$ confidence level, and
$n=2.886^{+0.169}_{-0.163}$ at $2\sigma$ confidence level. Then from
Eq.(\ref{HIH0}), we know
\begin{equation}
  \label{HIH0n}
  n^2\frac{H_I^2a_I^2}{H_0^2a_0^2}\ll1.
\end{equation}
With the choice of $a_0=1$, we may rewrite Eq.(\ref{OmegaqLess}) as
\begin{equation}
  \label{OmegaqLessa0}
  \Omega_{q0}<n^2\frac{a_I^2H_I^2}{a_0^2H_0^2}\Big(1-\frac{a_IH_I}{a_EH_E}\Big)^{-2}.
\end{equation}
At the same time, we know
\[
  a_IH_I\ll a_EH_E.
\]
Then, from Eq.(\ref{OmegaqLessa0}), we can conclude
\begin{equation}
  \label{OmegaqLL1}
  \Omega_{q0}\ll1.
\end{equation}
This is also unacceptable. Then, using Eq.(\ref{etaInflDelta}), the
zero point of $\eta$ is removed and the divergence of $\rho_q$ is
avoided, but we find that with Eq.(\ref{etaInflDelta}), NADE can not
become dominated at the present time.

To summary, in the note, we show that Eq.(\ref{etaInflation}) that
is widely adopted in the inflation models implies the existence of
the zero point of $\eta$ which subsequently indicates the divergence
of $\rho_q$. So Eq.(\ref{etaInflation}) is incompatible with the
NADE models. It seems that the incompatibility can be removed easily
if Eq.(\ref{etaInflation}) is replaced by Eq.(\ref{etaInflDelta}).
However, we find that Eq.(\ref{etaInflDelta}) indicates that
$\Omega_{q0}\ll1$. Then we think that the inflation scenario may be
incompatible with the NADE model. Here we note that the inflation
scenario and the NADE model are very successful in their own fields.
Although it has been shown that there exists the contradiction
between them, we do not think that either the inflation scenario or
the NADE model is incorrect. There may exists some unknown way out
of the incompatibility.

\vspace{5mm}

\noindent {\bf Acknowledgments} This work is supported in part by
the Natural Science Foundation of the Northwest University of China
under Grant No. 09NW27, and the National Natural Science Foundation
of China under Grant No. 10875060.

\end{document}